\documentclass[letter]{article}

\usepackage{amsmath}
\usepackage{amsfonts}
\usepackage{amssymb}
\usepackage{epsfig}
\usepackage{graphicx}

\begin{document}

\title{Non inertial dynamics and holonomy on the ellipsoid}
\author{Jos\'{e} A. Vallejo \\
Departament de Matem\`{a}tica Aplicada IV\\
Universitat Polit\`{e}cnica de Catalunya\\
Escola Polit\`{e}cnica Superior de Castelldefels\\
Av. Canal Ol\'{\i}mpic s/n\\
08860 Castelldefels (Spain)\\
e-mail: jvallejo@ma4.upc.edu}
\date{March 23 2006}
\maketitle

\begin{abstract}
Traditionally, the discussion about the geometrical interpretation of
inertial forces is reserved for General Relativity handbooks. In these notes
an analysis of the effect of such forces in a classical (newtonian) context
is made, as well as a study of the relation between the precession of the
Foucault pendulum and the holonomy on a surface, including some critical
remarks about the common statement \textquotedblleft precession of Foucault
pendulum equals holonomy on the sphere\textquotedblright.
\end{abstract}

\section{Introduction}

These notes intend to analyze a classical problem in Physics, that of
inertial forces, with the methods of elementary differential geometry of
curves and surfaces. Inertial forces appear in the study of the dynamics of
non inertial reference systems, and the problem they pose is that of their
interpretation in classical (Newtonian) terms.

In Newton's theory, the inertial forces are a consequence of the action of
absolute space on physical bodies; however, Mach and Einstein eradicated the
privileged r\^{o}le of this absolute space as a theoretical element, and
this fact led them to a reinterpretation of inertial forces (see the
discussion in \cite{Rin 77}, pgs $10-11$). Following them, we could say that
Newtonian conceptions have a strong unilateral character (in Einstein's
words \textquotedblleft In the first place, it is contrary to the mode of
thinking in science to conceive a thing (absolute space) which acts itself,
but which cannot be acted upon. This is the reason why E. Mach was led to
make the attempt to eliminate space as an active cause in the system of
mechanics\textquotedblright. See \cite{Ein 56} pages $55-56$) and this
reflects on the fact that, for inertial forces, the (third) Newtonian law of
action-reaction is not satisfied. To summarize, the conceptual difficulty
regarding inertial forces is that they are not real forces, with origin in
some known physical interaction (this motivates the term \textquotedblleft
fictitious forces\textquotedblright, sometimes used to describe them).

The solution offered by Einstein is well known: he proposed a geometrical
interpretation of accelerations (thus also of forces) expressed in his
famous \textquotedblleft equivalence principle\textquotedblright, according
to which the inertial and gravitational forces can be absorbed into a non
Euclidean structure of spacetime (see \cite{Lan 87} $\S 81$, \cite{Wei 72}
Section $1.3$). However, much care must be taken when doing this translation
from inertial forces to geometrical (non euclidean) structures. For example,
it is a common place (see \cite{Mar 92}, \cite{Sha 88}) that Newton
equations in the non inertial reference system associated to the Earth,
applied to the Foucault pendulum motion are equivalent to the holonomy on
the sphere, a typical phenomenon of non Euclidean geometry. However,
sometimes this fact is presented as something miraculous which calls for an
explanation: geometry determines dynamics. But, how it is possible that the
geometry of the surface on which it moves influence the plane of
oscillations of the pendulum? We will argue that this kind of questions must
be taken \textquotedblleft with a grain of salt\textquotedblright: the
holonomy on the sphere is \emph{not} responsible of the pendulum precession,
rather (as a result of some \emph{physical} approximations), the plane of
oscillations of the pendulum is a parallel vector field on the sphere, in
such a way that a physical effect, its total precession in a day, coincides
with the holonomy (we could say that both effects agree \emph{at first order}%
). But if we modify slightly the model to study the motion of the pendulum,
by considering the more realistic case of an ellipsoid, this coincidence is
no longer true.

In brief, the whole plan of this paper is as follows. The physical analysis
of the Foucault pendulum fits into a wider class of problems, called
dynamics in non inertial systems; in Section \ref{sec2} we present a r\'{e}%
sum\'{e} of the general ideas concerning it. Next follows a reminder of the
basic notions of Differential Geometry that will be used and, finally, we
will discuss how to apply these results to the Foucault pendulum.

Let us mention that there exist in the literature studies with a similar
spirit. Specifically, the excellent paper of J. Oprea (see \cite{Opr 95})
has been a stimulus for this work. However, the development that can be
found there can be somewhat confusing from a physical point of view, because
it makes an explicit use of the inertial \textquotedblleft
forces\textquotedblright\ and, by contrast, \emph{it does not account for
real forces as the string tension}, so it can be hardly seen as a correct
physical analysis. Indeed, this influences the crucial idea that the field
of pendulum oscillations is parallel, as if the tension is taken into
account it is no longer possible to directly conclude that \textquotedblleft
on the plane of oscillations of the pendulum there are no tangential
forces\textquotedblright\ (cfr. \cite{Opr 95}, pg $521$).

We will make use only of the machinery of the basic Differential Geometry of
curves and surfaces, in particular, no manifold theory or Riemannian
geometry will be needed. Suitable references are \cite{DoC 77}, \cite{Gra 98}
or \cite{Dub 98}.

\section{Non inertial Newtonian dynamics\label{sec2}}

A more or less standard statement of Newton's Laws could be:

\begin{quotation}
LAW I: Free (not interacting) particles follow rectilinear uniform motion
(including the limit case of being at rest).

LAW II: The acceleration of a body is inversely proportional to its mass,
and directly proportional to the external force which acts on it:%
\begin{equation}
\overrightarrow{F}=m\cdot\overrightarrow{a}.   \label{e1}
\end{equation}

LAW III: If a body $A$ acts upon another body $B$ with a certain force, the
body $B$ will act upon $A$ with an equal in magnitude but opposed force
along the line joining the bodies.
\end{quotation}

Newton himself states that his laws are applicable just in inertial frames.
A frame is basically a collection of physical and geometrical objects which
allows us to assign a set of numerical coordinates to any event (Einstein
described them intuitively as a coordinate system or reference and a clock
in each point of space). Newton considered that such a frame with an
absolute character exists (a very interesting discussion is given in \cite%
{Rin 77}, pgs. $2$ and $14-15$) and defined the inertial frames as those
moving with a uniform rectilinear velocity with respect to the absolute
frame, an idea already present in Galileo.

From an operational point of view, the inertial frames can be distinguished
with the aid of the first law, but what happens with non inertial frames
(NIF) as in the case of two observers with uniform rotational relative
motion?. In these systems the preceding laws are no longer valid. The
classical approach tries to continue making use of the inertial frames
machinery and \emph{forces the validity of the second law, at the cost of
introducing the so called inertial forces}. Thus, in a non inertial frame
the second law is not (\ref{e1}) but

\begin{quotation}
LAW II': The acceleration of a body is inversely proportional to its mass,
and directly proportional to the total force which acts on it, including
external (real) and inertial forces, $\overrightarrow{F}_{ext}$ and $%
\overrightarrow {F}_{in}$:%
\begin{equation}
\overrightarrow{F}_{in}=m\cdot\overrightarrow{a}-\overrightarrow{F}_{ext}. 
\label{e2}
\end{equation}
\end{quotation}

\textbf{Example 1:} In the particular case of a frame with uniform
rotational motion with respect to another fixed one (this one considered as
inertial), $\overrightarrow{F}_{in}$ takes the form of the Coriolis
\textquotedblleft force\textquotedblright:%
\begin{equation*}
-\overrightarrow{\omega}\wedge(\overrightarrow{\omega}\wedge\overrightarrow {%
r})-2\overrightarrow{\omega}\wedge\overrightarrow{v}_{rot}, 
\end{equation*}
where $\overrightarrow{\omega}$ is the rotation velocity of the rotating
system with respect to the fixed one, $\overrightarrow{r}$ is the vector
joining the origins of both systems and $\overrightarrow{v}_{rot}$ is the
velocity with respect to the rotating system.

The appearance of the inertial \textquotedblleft forces\textquotedblright
can be seen as the consequence of making a change of coordinates, from a set
of Euclidean coordinates (in the example those of the fixed-inertial system)
for which the Christoffel symbols equal zero, to another set of arbitrary
ones without this property (see \cite{Kop 92}, pg. $113$). These terms
involving the $\Gamma_{jk}^{i}$ are precisely those generating $%
\overrightarrow{F}_{in}$. On the other hand, the Christoffel symbols are
intimately related to intrinsical magnitudes such as curvature, so it seems
plausible to ask if the effect of inertial forces can be understood in terms
of these intrinsical magnitudes. Intuitively, we could say that the idea is
to replace equation (\ref{e2}) by%
\begin{equation*}
\overrightarrow{F}_{in}=``geometrical\text{ }effect\textquotedblright. 
\end{equation*}

As we have remarked, for the Foucault pendulum it has been suggested that
the geometrical effect is the holonomy on the sphere. However, as we will
see, this must be carefully analyzed.

\section{Geodesic curvature and holonomy}

It is well known that in the case of curves in the Euclidean space the
Frenet frame gives an $\mathbb{R}^{3}$ basis (a reference) which is adapted
to the geometry of the curve. Now, we will work with a curve contained in a
regular surface and so we will consider a similar object, that is, an
adapted basis not only to the curve but also to the surface.

Let $S\subset\mathbb{R}^{3}$ be an oriented regular surface and $%
c:I\rightarrow S$ with $c:t\mapsto c(t)$ a curve on it. We will call an 
\emph{intrinsic Frenet frame} a reference along $c$, denoted $\left\{
E_{1},E_{2},E_{3}\right\} $, and defined as follows (the point in $\dot{c}$
denotes derivation with respect to $t$, and $N$ is the unitary vector in $%
\mathbb{R}^{3}$ normal to the surface. The symbol $\wedge$ denotes the cross
product in $\mathbb{R}^{3}$):%
\begin{equation*}
\left\{ 
\begin{array}{l}
E_{1}(t)=\frac{\dot{c}(t)}{\left\vert c(t)\right\vert } \\ 
\\ 
E_{2}(t)=N\left( c(t)\right) \wedge E_{1}(t) \\ 
\\ 
E_{3}(t)=N\left( c(t)\right) .%
\end{array}
\right. 
\end{equation*}

By construction, $\left\{ E_{1},E_{2}\right\} $ is an orthonormal basis of $%
T_{c(t)}S$, the tangent plane to $S$ at $c(t)$. If we represent by $%
\left\langle \cdot,\cdot\right\rangle $ the Euclidean scalar product in $%
\mathbb{R}^{3}$ and by $\frac{DV}{dt}(t)$ the covariant derivative of the
vector field $V(t)$ on $S$ (that is, the projection of $\dot{V}(t)$ on $%
T_{c(t)}S$ with respect to the Euclidean scalar product of the ambient
space), then it is clear that%
\begin{equation*}
\left\langle E_{1}(t),\frac{DE_{1}}{dt}(t)\right\rangle =\left\langle
E_{1}(t),\dot{E}_{1}(t)\right\rangle =0, 
\end{equation*}
so for each $t\in I,$ $\frac{DE_{1}}{dt}(t)$ must be colinear with $E_{2}(t)$%
; thus%
\begin{equation*}
\frac{DE_{1}}{dt}(t)=\left\langle E_{2}(t),\frac{DE_{1}}{dt}(t)\right\rangle
E_{2}(t). 
\end{equation*}

The geodesic curvature of $c$\ at $t$ is the proportionality factor between $%
\frac{DE_{1}}{dt}(t)$ and $E_{2}(t)$. More precisely, let $S\subset \mathbb{R%
}^{3}$ be an oriented regular surface and $c:I\rightarrow S$ a curve on it.
Its geodesic curvature is the function%
\begin{equation}
\kappa_{g}(t)=\frac{1}{\left\vert \dot{c}(t)\right\vert }\left\langle
E_{2}(t),\frac{DE_{1}}{dt}(t)\right\rangle .   \label{e0}
\end{equation}
Geometrically, $\kappa_{g}$ expresses the angle between the acceleration of
the curve and the normal to the surface. When this angle is $0$, the
acceleration is tangent to the surface and the curve is said to be a \emph{%
geodesic}. It turns out that, for the sphere, the geodesics are given by the
intersection with planes passing through the origin, that is, the great
circles.

Let $W(t)$ be a vector field on $S$ parallel along $c$, that is a vector
field such that $\frac{DW}{dt}(t)=0$ for all $t\in I$. Let us call $\phi$
the angle it defines, at each point, with the tangent to the curve $c$:%
\begin{equation*}
\phi(t)=\measuredangle\left( \dot{c}(t),W(t)\right) . 
\end{equation*}
Under these assumptions, we have a basic result (cfr. \cite{DoC 77}, pg. 252
or \cite{One 97}, pg 338) telling us that%
\begin{equation*}
\dot{\phi}(t)=-\kappa_{g}(t)\left\vert \dot{c}(t)\right\vert , 
\end{equation*}
that is, the variation of the angle $\phi$ is a characteristic of the curve $%
c$. This expression, is particularly suited to the practical computation of
the parallel transport. Let $t_{1},t_{2}\in I$ be two instants, then the
parallel transport along $c$ between $t_{1}$ and $t_{2}$ mapping, denoted $%
\tau(c)_{t_{1}}^{t_{2}}:T_{c(t_{1})}S\rightarrow T_{c(t_{2})}S$, carries $%
v\in T_{c(t_{1})}S$ to%
\begin{equation*}
\tau(c)_{t_{1}}^{t_{2}}(v)=\left\vert v\right\vert \left(
E_{1}(t_{2})\cos\phi(t_{2})+E_{2}(t_{2})\sin\phi(t_{2})\right) , 
\end{equation*}
where%
\begin{equation*}
\phi(t_{2})=\phi(t_{1})-\int\nolimits_{t_{1}}^{t_{2}}\kappa_{g}(t)\left\vert 
\dot{c}(t)\right\vert dt. 
\end{equation*}
According to this formula, if we have a closed curve on $S$ (parametrized in 
$[0,2\pi]$ without loss of generality) with $c(0)=c(2\pi)=p$, when parallel
transporting a tangent vector $v$ in $p$, making a complete loop around $c$,
we will not recover $v$ but a vector defining a certain angle with it; this
phenomenon is known as the \emph{holonomy}.

More precisely, with the preceding conditions the \emph{angle of holonomy of 
}$c$\emph{\ in }$p$ is defined as the angle between $v$ and $%
\tau(c)_{0}^{2\pi}(v)$,%
\begin{equation*}
\hbar=\measuredangle\left( v,\tau(c)_{0}^{2\pi}(v)\right) . 
\end{equation*}

\textbf{Example 2}: Let us compute the holonomy of a parallel on the sphere $%
S^{2}$ (of unit radius). Consider the geographical parametrization, given by%
\begin{equation*}
\begin{array}{l}
X:[-\frac{\pi}{2},\frac{\pi}{2}]\times\lbrack0,2\pi]\rightarrow S^{2} \\ 
(u,v)\mapsto X(u,v)=(\cos u\cos v,\cos u\sin v,\sin u),%
\end{array}
\end{equation*}
and the parallel of latitude $\theta_{0}$, $c(t)=(\cos\theta_{0}\cos
t,\cos\theta_{0}\sin t,\sin\theta_{0})$; it is easy to see that its geodesic
curvature is constant with value (note that the sign depends on the
orientation of $S^{2}$ chosen)%
\begin{equation*}
\kappa_{g}(t)=-\tan\theta_{0}, 
\end{equation*}
while%
\begin{equation*}
\left\vert \dot{c}(t)\right\vert =\cos\theta_{0}, 
\end{equation*}
so%
\begin{equation}
\hbar=\int\nolimits_{0}^{2\pi}\kappa_{g}(t)\left\vert \dot{c}(t)\right\vert
dt=\int\nolimits_{0}^{2\pi}\sin\theta_{0}dt=2\pi\sin\theta_{0}.   \label{e3}
\end{equation}

\textbf{Example 3}: A model for Earth's surface more realistic than a sphere
is an ellipsoid of revolution $E_{a,b}$ (with distinct semiaxes $a$ and $b$%
). It can also be parametrized geographically:%
\begin{equation*}
\begin{array}{l}
X:[-\frac{\pi}{2},\frac{\pi}{2}]\times\lbrack0,2\pi]\rightarrow E_{a,b} \\ 
(u,v)\mapsto X(u,v)=(a\cos u\cos v,a\cos u\sin v,b\sin u),%
\end{array}
\end{equation*}
and a computation as in Example $2$ gives for the parallel on the ellipsoid $%
c(t)=(a\cos\theta_{0}\cos t,a\cos\theta_{0}\sin t,b\sin\theta_{0})$ the
results 
\begin{equation*}
\kappa_{g}(t)=-\frac{\tan\theta_{0}}{\sqrt{a^{2}\sin^{2}\theta_{0}+b^{2}%
\cos^{2}\theta_{0}}}
\end{equation*}
and%
\begin{equation}
\hbar=\frac{2\pi a\sin\theta_{0}}{\sqrt{a^{2}\sin^{2}\theta_{0}+b^{2}\cos
^{2}\theta_{0}}}.   \label{e4}
\end{equation}
Of course, when $a=b$ these coincide with the corresponding to the sphere.

\section{Parallel vector fields on the sphere and the ellipsoid\label{Sphere}%
}

Consider a curve $c:I\rightarrow S\subset\mathbb{R}^{3}$ and $X:U\subset 
\mathbb{R}^{2}\rightarrow S$ a parametrization of the surface $S$, so $%
c(t)=X(u(t),v(t))$. Let $V$ be a vector field along $c$ such that in the
reference $\left\{ X_{u}(t),X_{v}(t)\right\} $ of $T_{c(t)}S$ has
coordinates $\left( V^{1}(t),V^{2}(t)\right) $, that is,%
\begin{equation}
V(t)=V^{1}(t)X_{u}(t)+V^{2}(t)X_{v}(t)\in T_{c(t)}S.   \label{e5}
\end{equation}
For a vector field with that expression, it is known that the covariant
derivative along $c$ is given by%
\begin{align*}
\frac{DV}{dt} & =\left( \dot{V}^{1}+\Gamma_{11}^{1}V^{1}\dot{u}+\Gamma
_{12}^{1}V^{1}\dot{v}+\Gamma_{12}^{1}V^{2}\dot{u}+\Gamma_{22}^{1}V^{2}\dot {v%
}\right) X_{u} \\
& +\left( \dot{V}^{2}+\Gamma_{11}^{2}V^{1}\dot{u}+\Gamma_{12}^{2}V^{1}\dot{v}%
+\Gamma_{12}^{2}V^{2}\dot{u}+\Gamma_{22}^{2}V^{2}\dot{v}\right) X_{v}.
\end{align*}
In particular, if we consider $S=S^{2}$ (the unit sphere), $X$ the
geographical parametrization of Example $2$ and $c(t)$ the parallel of
latitude $\theta_{0}$, we get%
\begin{equation*}
\begin{array}{ll}
u=\theta_{0}, & \dot{u}=0 \\ 
v=t & \dot{v}=1.%
\end{array}
\end{equation*}
Also, for the Christoffel symbols:%
\begin{equation*}
\begin{array}{lll}
\begin{array}{l}
\Gamma_{11}^{1}=0, \\ 
\end{array}
& 
\begin{array}{l}
\Gamma_{12}^{1}=0, \\ 
\end{array}
& 
\begin{array}{l}
\Gamma_{22}^{1}=\sin u\cos u \\ 
\end{array}
\\ 
\Gamma_{11}^{2}=0, & \Gamma_{12}^{2}=-\tan u, & \Gamma_{22}^{2}=0,%
\end{array}
\end{equation*}
so%
\begin{equation*}
\frac{DV}{dt}(t)=\left( \dot{V}^{1}(t)+\sin\theta_{0}\cos\theta_{0}V^{2}(t)%
\right) X_{u}+\left( \dot{V}^{2}(t)-\tan\theta_{0}V^{1}(t)\right) X_{v}. 
\end{equation*}

According to that expression, a vector field along a parallel of latitude $%
\theta_{0}$ on the sphere is parallel if and only if its components with
respect to the basis $\left\{ X_{u}(t),X_{v}(t)\right\} $ (recall that $X$
is the geographic parametrization) verify%
\begin{equation}
\left\{ 
\begin{array}{l}
\dot{V}^{1}(t)+\sin\theta_{0}\cos\theta_{0}V^{2}(t)=0 \\ 
\\ 
\dot{V}^{2}(t)-\tan\theta_{0}V^{1}(t)=0.%
\end{array}
\right.   \label{e5b}
\end{equation}
If we write $\beta=\sin\theta_{0},x=V^{1},y=\cos\theta_{0}V^{2}$ this system
is equivalent to%
\begin{equation*}
\left\{ 
\begin{array}{l}
\frac{\dot{x}}{\beta}+y=0 \\ 
\\ 
\dot{y}-\beta x=0,%
\end{array}
\right. 
\end{equation*}
which has solution%
\begin{equation*}
\left\{ 
\begin{array}{l}
y(t)=B\sin\beta t+A\cos\beta t \\ 
\\ 
x(t)=-A\sin\beta t+B\cos\beta t.%
\end{array}
\right. 
\end{equation*}

Thus, the more general expression for a parallel vector field along the
curve $c(t)=(\cos\theta_{0}\cos t,\cos\theta_{0}\sin t,\sin\theta_{0})$ (in
the geographical parametrization) on the sphere $S^{2}$ is%
\begin{align}
V(t) & =\left( -A\sin\beta t+B\cos\beta t\right) X_{u}(t)  \label{e6} \\
& +\frac{1}{\cos\theta_{0}}\left( B\sin\beta t+A\cos\beta t\right) X_{v}(t).
\notag
\end{align}

Given the Christoffel symbols for the ellipsoid $E_{a,b}$:%
\begin{equation*}
\begin{array}{lll}
\begin{array}{l}
\Gamma_{11}^{1}=\frac{2\sin u\cos u(b^{2}-a^{2})}{a^{2}\sin^{2}u+b^{2}\cos
^{2}u}, \\ 
\end{array}
& 
\begin{array}{l}
\Gamma_{12}^{1}=0, \\ 
\end{array}
& 
\begin{array}{l}
\Gamma_{22}^{1}=\frac{a^{2}\sin u\cos u}{a^{2}\sin^{2}u+b^{2}\cos^{2}u} \\ 
\end{array}
\\ 
\Gamma_{11}^{2}=0, & \Gamma_{12}^{2}=-\tan u, & \Gamma_{22}^{2}=0,%
\end{array}
\end{equation*}
a similar analysis shows that the more general expression for a parallel
vector field along the curve $c(t)=(\cos\theta_{0}\cos t,\cos\theta_{0}\sin
t,\sin\theta_{0})$ (in the geographical parametrization) on the ellipsoid is
given by a vector $V(t)=V^{1}(t)X_{u}(t)+V^{2}(t)X_{v}(t)$ whose components
satisfy a system of equations similar to (\ref{e5b}):%
\begin{equation}
\left\{ 
\begin{array}{l}
\dot{V}^{1}(t)+\frac{a^{2}\sin\theta_{0}\cos\theta_{0}}{a^{2}\sin^{2}%
\theta_{0}+b^{2}\cos^{2}\theta_{0}}V^{2}(t)=0 \\ 
\\ 
\cos\theta_{0}\dot{V}^{2}(t)-\sin\theta_{0}V^{1}(t)=0.%
\end{array}
\right.   \label{e6b}
\end{equation}

\section{Pendula, Earth rotation and holonomy}

To carry on our analysis, we begin with two basic \emph{assumptions}. The
first, that the Earth surface is spherical (and, as a consequence, the
gravitational field it creates). The second, that the \emph{Cartesian} frame
with origin the center of the Earth sphere is (approximately) inertial. In
order to do explicit computations we will consider the geographical
parametrization of the sphere, and we will choose a non inertial frame
associated to the parallel $u=\theta_{0}$ as the intrinsic Frenet frame,
that is: at each point $c(t=v)$ the \textquotedblleft$X$ axis%
\textquotedblright\ is the direction of $E_{1}(t)$, the \textquotedblleft$Y$
axis\textquotedblright\ is the direction of $E_{2}(t)$ and the
\textquotedblleft$Z$ axis\textquotedblright\ is the direction of $E_{3}(t)$.

\medskip

\begin{figure}[h]
\centerline{ \epsfig{file=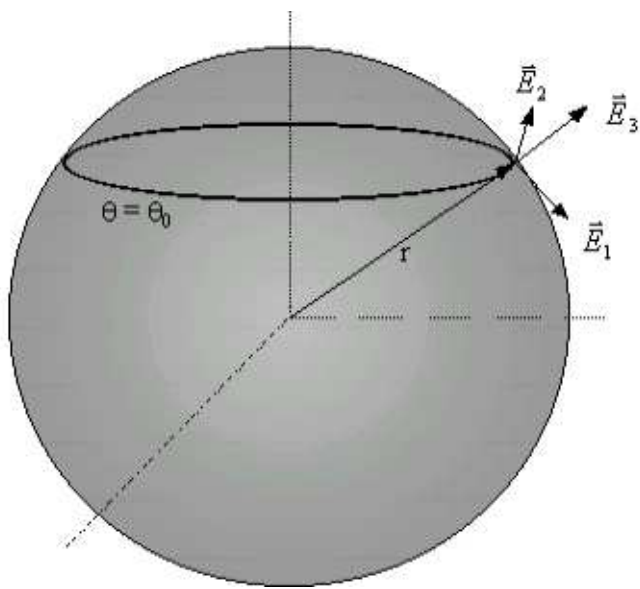, width=10cm} }
\caption{The Frenet frame.}
\label{fig1}
\end{figure}

\medskip

Imagine the daily motion of the Foucault pendulum as the result of
displacing the pendulum along a parallel on an Earth which stand still.
Then, this non inertial frame will be rotating jointly with the pendulum.
For the sake of simplicity, let us suppose that the mass of the bob is $m=1$
and that we set the time scale in such a way that $1$ day equals $2\pi$
units of time, that is, the module of the angular velocity of the pendulum
is $\omega=1$. If $\overrightarrow{r}$ is the vector giving the position of
the origin of the rotating system in the Cartesian one and $\overrightarrow{v%
}_{f},\overrightarrow{a}_{f},\overrightarrow{v}_{rot},\overrightarrow{a}%
_{rot}$ are the velocity and acceleration (respectively) with respect to the
fixed Cartesian axes and the rotating system then, for an observer in this
rotating system the effective force on the pendulum is (see \cite{Cho 95},
Sec. $10.4$ for the following development)%
\begin{equation*}
\overrightarrow{F_{ef}}=\overrightarrow{a_{rot}}=\overrightarrow{a_{f}}-%
\overrightarrow{\omega}\wedge(\overrightarrow{\omega}\wedge\overrightarrow {r%
})-2\overrightarrow{\omega}\wedge\overrightarrow{v_{rot}}. 
\end{equation*}

Thus, the equation of motion becomes 
\begin{equation}
\overrightarrow{a_{rot}}=\overrightarrow{g}+\overrightarrow{T}-2%
\overrightarrow{\omega}\wedge\overrightarrow{v_{rot}},   \label{e7}
\end{equation}
where $\overrightarrow{T}$ is the string tension and $\overrightarrow{g}$ is
the gravitational force.

\textbf{Remark:} Note that the term $-\overrightarrow{\omega}\wedge (%
\overrightarrow{\omega}\wedge\overrightarrow{r})$ can be included in $%
\overrightarrow{g}$, as $\overrightarrow{g}$ is only defined through the
measures we make. Incidentally, this is the cause of the fact that the
direction of $\overrightarrow{g}$ actually differs from the local vertical.
However, we will assume that $\overrightarrow{g}$ is directed along $%
E_{3}(t) $.

Here comes our third assumption: if the displacements of the bob have little
amplitude, so the approximation $\sin\zeta\simeq\zeta$ is valid (where $\zeta
$ is the angle that the string determines with the vertical), and $l$ is the
string longitude, we have%
\begin{equation*}
\begin{array}{l}
T_{x}\simeq-Tx/l \\ 
T_{y}\simeq-Ty/l \\ 
T_{z}\simeq T%
\end{array}
\end{equation*}

\begin{figure}[h]
\centerline{ \epsfig{file=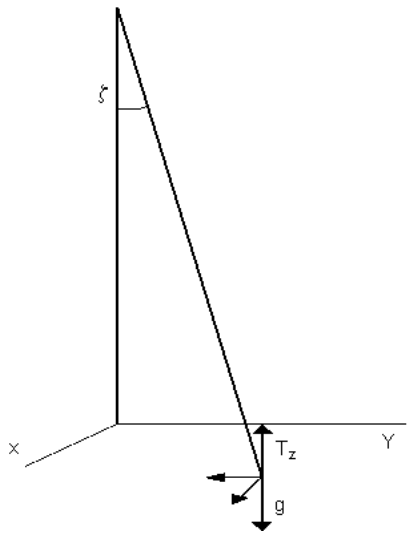, width=10cm} }
\caption{Forces on the pendulum.}
\label{fig2}
\end{figure}

\medskip

Also, by the symmetry of the gravitational field (we are supposing the Earth
spherical), 
\begin{equation*}
\begin{array}{ccc}
\left\{ 
\begin{array}{l}
g_{x}=0 \\ 
g_{y}=0 \\ 
g_{z}=-g%
\end{array}
\right. & \left\{ 
\begin{array}{l}
\omega_{x}=-\cos\theta_{0} \\ 
\omega_{y}=0 \\ 
\omega_{z}=\sin\theta_{0}%
\end{array}
\right. & \left\{ 
\begin{array}{l}
\left( \overrightarrow{v}_{rot}\right) _{x}=\dot{x} \\ 
\left( \overrightarrow{v}_{rot}\right) _{y}=\dot{y} \\ 
\left( \overrightarrow{v}_{rot}\right) _{z}=0.%
\end{array}
\right.%
\end{array}
\end{equation*}
Then, computing $\overrightarrow{\omega}\wedge\overrightarrow{r}$ we get 
\begin{equation*}
\left\{ 
\begin{array}{l}
\left( \overrightarrow{\omega}\wedge\overrightarrow{v}_{rot}\right) _{x}=-%
\dot{y}\sin\theta_{0} \\ 
\\ 
\left( \overrightarrow{\omega}\wedge\overrightarrow{v}_{rot}\right) _{y}=%
\dot{x}\sin\theta_{0} \\ 
\\ 
\left( \overrightarrow{\omega}\wedge\overrightarrow{v}_{rot}\right) _{z}=-%
\dot{y}\cos\theta_{0},%
\end{array}
\right. 
\end{equation*}
and the equations we are interested in are:%
\begin{equation*}
\left\{ 
\begin{array}{l}
\left( \overrightarrow{a}_{rot}\right) _{x}=\ddot{x}=-Tx/l+2\dot{y}%
\sin\theta_{0} \\ 
\\ 
\left( \overrightarrow{a}_{rot}\right) _{y}=\ddot{y}=-Ty/l-2\dot{x}%
\sin\theta_{0}.%
\end{array}
\right. 
\end{equation*}

For displacements of little amplitude, the tension and the weight are
approximately equal $T\simeq g$. If we set 
\begin{equation*}
\alpha^{2}=\frac{T}{l}=\frac{g}{l}, 
\end{equation*}
and 
\begin{equation*}
\beta=\sin\theta_{0}, 
\end{equation*}
we arrive at the (approximate) equations 
\begin{equation*}
\left\{ 
\begin{array}{l}
\ddot{x}+\alpha^{2}x=2\beta\dot{y} \\ 
\ddot{y}+\alpha^{2}y=-2\beta\dot{x},%
\end{array}
\right. 
\end{equation*}
a system of coupled second order ordinary differential equations that are
the mathematical model for the Foucault pendulum within the approximation of
little amplitude displacements. It can be solved adding to the first
equation the second multiplied by $i$, so (writing $q=x+iy)$ we get%
\begin{equation*}
\ddot{q}+2i\beta\dot{q}+\alpha^{2}q=0, 
\end{equation*}
with solution%
\begin{equation*}
q\left( t\right) \simeq\exp(-i\beta t)(A\exp(-t\sqrt{\beta^{2}-\alpha^{2}}%
)+B\exp(-t\sqrt{-\beta^{2}-\alpha^{2}})), 
\end{equation*}
$\alpha$ being the pendulum's pulse of oscillation. Of course, this is much
greater than the rotational velocity of the Earth, so $\alpha\gg\beta$ and%
\begin{equation*}
q(t)=\exp(-i\beta t)(A\exp(i\alpha t)+B\exp(-i\alpha t)), 
\end{equation*}
or, if we denote by $q_{0}(t)$ the solution corresponding to the absence of
rotation (a static pendulum),%
\begin{equation*}
q(t)\simeq q_{0}(t)\exp(-i\beta t). 
\end{equation*}
Equating real and imaginary parts:%
\begin{equation}
\left( 
\begin{array}{c}
x(t) \\ 
y(t)%
\end{array}
\right) =\left( 
\begin{array}{cc}
\cos(\beta t) & \sin(\beta t) \\ 
-\sin(\beta t) & \cos(\beta t)%
\end{array}
\right) \left( 
\begin{array}{c}
x_{0}(t) \\ 
y_{0}(t)%
\end{array}
\right)   \label{e8}
\end{equation}
where $\left( x_{0}(t),y_{0}(t)\right) $ is the \textquotedblleft
static\textquotedblright\ solution for the pendulum in $c(0)$. As is well
known, this oscillation would always take place in a plane determined by the
initial conditions; what (\ref{e8}) is telling us is that this plane
precesses when the pendulum is displaced along the parallel $c(t)$, as was
noted by Foucault.

Let us check that \emph{the dynamics of the problem determines a vector
field (associated to the pendulum oscillations) which is parallel along }$c$%
\emph{, so the total angle accumulated after a complete loop will coincide
with the holonomy of }$c$, given by (\ref{e3}).

As we have already mentioned, the initial oscillation defines a unitary
vector through the intersection of the plane in which it is contained (in $%
\mathbb{R}^{3}$) and the tangent plane to the sphere in $c(0)$, let us call
it%
\begin{equation*}
\left( 
\begin{array}{c}
A \\ 
B%
\end{array}
\right) \in T_{c(0)}S^{2}, 
\end{equation*}
and what we have is that $\left( x_{0}(t),y_{0}(t)\right) $ is parallel to $%
(A,B)$:%
\begin{equation*}
\left( 
\begin{array}{c}
x_{0}(t) \\ 
y_{0}(t)%
\end{array}
\right) \propto\left( 
\begin{array}{c}
A \\ 
B%
\end{array}
\right) . 
\end{equation*}

Now, the vector field associated to the oscillations of the pendulum
(defined by the intersection of the oscillation plane with the tangent plane
at each point) is%
\begin{equation*}
\begin{array}{l}
V:[0,2\pi]\rightarrow TS \\ 
t\mapsto V(t)=x(t)E_{1}(t)+y(t)E_{2}(t),%
\end{array}
\end{equation*}
where, from (\ref{e8}),%
\begin{equation*}
\left\{ 
\begin{array}{l}
x(t)=B\sin\beta t+A\cos\beta t \\ 
\\ 
y(t)=-A\sin\beta t+B\cos\beta t,%
\end{array}
\right. 
\end{equation*}
so, as time goes by, the plane of oscillation actually precesses with
respect to the plane of the initial oscillations.

Let us note that in the geographical parametrization the following relations
hold:%
\begin{equation*}
\begin{array}{l}
E_{1}(t)=\frac{X_{t}(\theta_{0},t)}{\left\Vert
X_{t}(\theta_{0},t)\right\Vert }=\frac{X_{v}(t)}{\cos\theta_{0}} \\ 
\\ 
E_{2}(t)=X_{u}(t),%
\end{array}
\end{equation*}
so we can write%
\begin{align}
V(t) & =(B\sin\beta t+A\cos\beta t)\frac{1}{\cos\theta_{0}}X_{v}(t)
\label{e9} \\
& +(-A\sin\beta t+B\cos\beta t)X_{u}(t).  \notag
\end{align}
Comparing with (\ref{e6}), we see that the vector field of the oscillations
is parallel transported, and the precession it experiments after a day (a
whole loop around $c$) is $2\pi\sin\theta_{0}$, according to (\ref{e3}).

\textbf{Remark:} As a curious observation, at the equator we would not find
any holonomy. From a geometrical point of view this is a consequence of the
fact that the equator is the only one among the parallels of the sphere that
is a geodesic (a great circle) and, consequently, has $\kappa_{g}(t)=0$
(recall the comments following (\ref{e0})).

\section{The ellipsoidal Earth}

The purpose of this Section is to analyze what happens when we change the
sphere model of the Earth. In order to understand it, we recall some
elementary facts about Geodesy.

The theoretical shape of the Earth is described by means of its normal
equipotential surface, which coincides with the mean level sea surface (in
firm land, the surface of a water channel) and is called the \emph{geoid}.
The geoid includes the variations in gravitational potential due, among
other factors, to the non homogeneity of the crust; this is the fundamental
reason that makes the description of the geoid only approximate.

The simplest approximation considers the geoid as a sphere (the ideas
introduced by holonomy are based on this assumption, as we have seen), but
another -much more precise- one is that of an ellipsoid of revolution.
Indeed, conventionally, the gravimetric reductions of experimental data are
carried on to a so called \emph{reference ellipsoid} defined by the
experimental values founded for some parameters (such as the equatorial
radius of the Earth $R$, its total mass $M$ or the flattening coefficient $f$%
) and \emph{the requirement that the ellipsoid surface be an equipotential
surface}. That is: starting from experimental data, we construct a
mathematical model for the gravitational potential of the Earth, in such a
way that the reference ellipsoid is an equipotential surface. The precise
form of this potential is not of interest for us (see \cite{Men 90} or \cite%
{Tor 91} for details), but what we will need is to retain the fact that in
the ellipsoidal model the gravitational field $\vec{g}$ is still normal at
each point of the surface. This will be of fundamental importance in what
follows.

Let us return for a moment to the computations for the sphere case in
Section \ref{Sphere}. We can see that the construction of the Frenet frame $%
\{E_{1}(t),E_{2}(t),E_{3}(t)\}$ is identical in the case of the ellipsoid,
as it is equation of motion (\ref{e7}) and the approximations following it
(which are based on the fact that $\vec{g}$ is directed along $E_{3}(t)$).
Thus, for the motion of the Foucault's pendulum along a parallel $c$ on the
ellipsoid we get again the equations%
\begin{equation*}
V(t)=(B\sin \beta t+A\cos \beta t)E_{1}(t)+(-A\sin \beta t+B\cos \beta
t)E_{2}(t).
\end{equation*}

As for the parallel $c(t)=X(\theta_{0},t)=(a\cos\theta_{0}\cos t,a\cos
\theta_{0}\sin t,b\sin t)$ on the ellipsoid we have%
\begin{align*}
E_{1}(t) & =\frac{X_{t}(\theta_{0},t)}{\left\Vert
X_{t}(\theta_{0},t)\right\Vert }=\frac{X_{v}(t)}{a\cos\theta_{0}}, \\
E_{1}(t) & =\frac{X_{u}(t)}{\sqrt{a^{2}\sin^{2}\theta_{0}+b^{2}\cos^{2}%
\theta_{0}}},
\end{align*}
we have, instead of (\ref{e9}):%
\begin{align*}
V(t) & =(B\sin\beta t+A\cos\beta t)\frac{1}{a\cos\theta_{0}}X_{v}(t) \\
& +(-A\sin\beta t+B\cos\beta t)\frac{1}{\sqrt{a^{2}\sin^{2}\theta_{0}+b^{2}%
\cos^{2}\theta_{0}}}X_{u}(t).
\end{align*}

This is to be compared with (\ref{e6b}), which gives the expression for a
parallel vector field along $c$ on the ellipsoid. What we see, is that the
components%
\begin{equation*}
V^{1}(t)=\frac{-A\sin\beta t+B\cos\beta t}{\sqrt{a^{2}\sin^{2}%
\theta_{0}+b^{2}\cos^{2}\theta_{0}}}
\end{equation*}
and%
\begin{equation*}
V^{2}(t)=\frac{B\sin\beta t+A\cos\beta t}{a\cos\theta_{0}}
\end{equation*}
do not satisfy (\ref{e6b}). For instance%
\begin{equation*}
tg\theta_{0}V^{1}(t)=\frac{\tan\theta_{0}}{\sqrt{a^{2}\sin^{2}%
\theta_{0}+b^{2}\cos^{2}\theta_{0}}}(-A\sin\beta t+B\cos\beta t) 
\end{equation*}
while (recall $\beta=\sin\theta_{0}$)%
\begin{equation*}
\dot{V}^{2}(t)=\frac{\tan\theta_{0}}{a}(-A\sin\beta t+B\cos\beta t), 
\end{equation*}
so cos$\theta_{0}\dot{V}^{2}(t)-\sin\theta_{0}V^{1}(t)\neq0$. However, we
see that the equation is true when $a=b$, as it should be.

To summarize: for the ellipsoidal Earth the vector field associated with the
displacement of the Foucault's pendulum is no longer parallel, so even it
does not make sense to speak about its holonomy.

\section{Conclusions}

We have seen that the holonomy can not be considered as the responsible for
the Foucault pendulum precession, as changing slightly the spherical model
of the Earth by an ellipsoid $E_{a,b}$ (which can be taken as
\textquotedblleft round\textquotedblright\ as wanted, just making $%
a\rightarrow b$) the numerical values of both effects no longer coincide
(indeed, it is nonsense to speak about holonomy in this case), this only
happens in the limit case $a=b$. From the analysis in the preceding
Sections, we can see explicitly that the holonomy only intervenes in the
precession of the pendulum as a first order effect: that the equation (\ref%
{e9}) describe a parallel vector field is true for the sphere after making a
set of \emph{physical} approximations, as considering only little amplitudes
or a spherically symmetric gravitational field. The more realistic situation
of an ellipsoidal Earth can be seen as a perturbation of the spherical case,
giving a first order term in which the holonomy appears, but not as the
definitive effect. Of course, it would be very interesting to study which
geometrical effects appear at higher orders.


\begin{thebibliography}{Gray 98}
\bibitem[Cho 95]{Cho 95} T. L. Chow: \emph{Classical Mechanics}. John Wiley
\& Sons, New York, 1995.

\bibitem[DoC 77]{DoC 77} M. P. do Carmo: \emph{Differential geometry of
curves and surfaces}. Prentice-Hall Inc. Englewood Cliffs (NJ), $1976$.

\bibitem[Dub 98]{Dub 98} B. A. Dubrovin, A. T. Fomenko, S. P. Novikov: \emph{%
Modern geometry, methods and applications. Part I: The geometry of surfaces,
transformation groups, and fields} ($2$nd ed). Graduate Texts in Mathematics 
$\mathbf{93}$. Springer-Verlag, New York, $1992$.

\bibitem[Ein 56]{Ein 56} A. Einstein: \emph{The meaning of relativity} ($5$%
th ed.). Princeton University Press; Princeton (NJ), 1956 ($4$th Princeton
paperback printing, 1974).

\bibitem[Gray 98]{Gra 98} A. Gray: \emph{Modern differential geometry of
curves and surfaces with Mathematica} ($2$nd. ed). CRC Press, Boca Raton
(FL), $1998$.

\bibitem[Kop 92]{Kop 92} W. Kopczynski, A. Trautman: \emph{Spacetime and
gravitation}. John Wiley \& Sons, Chichester; Polish Scientific Publishers,
Warsaw, $1992$.

\bibitem[Lan 87]{Lan 87} L. D. Landau, E. M. Lifshitz: \emph{The classical
theory of fields} ($2$nd. rev. ed). Course of Theoretical Physics, Vol. $%
\mathbf{2}$. Pergamon Press, Oxford; Addison-Wesley Publishing Co. Inc.
Reading (MASS), $1962$.

\bibitem[Mar 92]{Mar 92} J. E. Marsden: \emph{Lectures on mechanics}. London
Mathematical Society Lecture Note Series $\mathbf{174}$. Cambridge
University Press, Cambridge, $1992$.

\bibitem[Men 90]{Men 90} W. Menke, D. Abbott: \emph{Geophysical Theory}.
Columbia University Press, New York, $1990$.

\bibitem[One 97]{One 97} B. O'Neill: \emph{Elementary Differential Geometry}%
. Academic Press, San Diego (CA), 1997.

\bibitem[Opr 95]{Opr 95} J. Oprea: \emph{Geometry and the Foucault pendulum}%
. Amer. Math. Monthly $\mathbf{102}$, $6$. June-July ($1995$), $515-522$.

\bibitem[Rin 77]{Rin 77} W. Rindler: \emph{Essential Relativity} ($2$nd.
rev. ed). Springer Verlag, New York, $1977$.

\bibitem[Sha 88]{Sha 88} A. Shapere, F. Wilczek (eds.): \emph{Geometric
phases in physics}. Advanced Series in Mathematical Physics $\mathbf{5}$.
World Scientific Publishing Co. Inc. Teaneck (NJ), $1989$.

\bibitem[Tor 91]{Tor 91} W. Torge: \emph{Geodesy} ($2$nd. ed.). Walter de
Gruyter, Berlin, $1991$.

\bibitem[Wei 72]{Wei 72} S. Weinberg: \emph{Gravitation and Cosmology}. John
Wiley \& Sons, New York, $1972$.
\end{thebibliography}
\end{document}